\documentclass[prb,reprint,a4paper,showpacs,citeautoscript,floatfix]{revtex4-1}

\usepackage{amsfonts,amsmath,amssymb}
\usepackage{mathptmx}
\usepackage{graphicx}
\usepackage{bm}\let\vec\bm
\usepackage{mathrsfs}
\usepackage[
	colorlinks=true,linkcolor=blue,citecolor=blue,filecolor=blue,urlcolor=blue,
	breaklinks=true,pagebackref=false,pdfstartview=FitH,bookmarksopen=false,
]{hyperref}

\begin{document}

\title{Tunneling spectra of strongly coupled superconductors: Role of dimensionality}

\author{C. Berthod}
\affiliation{DPMC-MaNEP, Universit\'e de Gen\`eve, 24 quai Ernest-Ansermet, 1211
Gen\`eve 4, Switzerland}

\date{\today}

\begin{abstract}

We investigate numerically the signatures of collective modes in the tunneling
spectra of superconductors. The larger strength of the signatures observed in
the high-$T_c$ superconductors, as compared to classical low-$T_c$ materials, is
explained by the low dimensionality of these layered compounds. We also show
that the strong-coupling structures are dips (zeros in the $d^2I/dV^2$ spectrum)
in $d$-wave superconductors, rather than the steps (peaks in $d^2I/dV^2$)
observed in classical $s$-wave superconductors. Finally we question the
usefulness of effective density of states  models for the analysis of tunneling
data in $d$-wave superconductors.

\end{abstract}

\maketitle

\section{Introduction}

Many experiments have shown that the electrons in cuprate high-$T_c$
superconductors (HTS) are significantly renormalized by the interaction with
collective modes. This renormalization appears in photoemission measurements as
velocity changes in the quasi-particle dispersion (the ``kinks'') accompanied by
a drop of the quasi-particle life-time \cite{Damascelli-2003, Campuzano-2004}.
In tunneling, the renormalization shows up as a depression, or ``dip'', in the
$dI/dV$ curve with the associated nearby accumulation of spectral weight (the
``hump'') \cite{Fischer-2007}. Similar signatures observed by tunneling
spectroscopy in classical superconductors were successfully explained by the
strong-coupling theory of superconductivity \cite{Eliashberg-1960,
Schrieffer-1964b, Carbotte-1990}. There are, however, two striking differences
between the structures observed in the cuprates and in low-$T_c$ metals such as
Pb or Hg. The dip in the cuprates is electron-hole asymmetric, being strongest
at negative bias, while no such asymmetry is seen in lead. The electron-hole
asymmetry of the dip is due to the electron-hole asymmetry of the underlying
electronic density of states (DOS) \cite{Eschrig-2000, Levy-2008}. Second, while
the structures are subtle in low-$T_c$ materials---they induce a change smaller
than 5\% in the tunneling spectrum---the cuprate dip is generally a strong
effect which, for instance, can reach $20\%$ in optimally doped
Bi$_2$Sr$_2$Ca$_2$Cu$_3$O$_{10+\delta}$ (Bi-2223) \cite{Kugler-2006}. It is
tempting to attribute this difference of intensities to a difference in the
overall coupling strength, as suggested by the largely different $T_c$ values.
However, a comparison of the effective masses indicates that the couplings are
not very different in Pb where \cite{Swihart-1965} $m^*/m=2.1$ and in the
Bi-based cuprates where $m^*/m$ varies between $1.5$ and $3$ as a function of
doping \cite{Johnson-2001, Hwang-2007}. Here we show that the large magnitude of
the dip feature results from the low dimensionality of the materials and the
associated singularities in the electronic DOS.

Tunneling experiments in strongly coupled classical superconductors have been
interpreted using a formalism \cite{Schrieffer-1963} that neglects the momentum
dependence of the Eliashberg functions and of the tunneling matrix element, and
further assumes that the normal-state DOS $N_0(\omega)$ is constant over the
energy range of interest. The tunneling conductance, then, only depends on the
gap function $\Delta(\omega)$, whose energy variation reflects the singularities
of the phonon spectrum \cite{Scalapino-1964, McMillan-1965}. The dimensionality
of the materials does not enter in this formalism. The effect of a non-constant
$N_0(\omega)$ on the gap function has been discussed in the context of the A15
compounds \cite{Pickett-1980}. However the direct effect of a rapidly varying
$N_0(\omega)$ on the tunneling conductance became apparent only recently in the
high-$T_c$ compounds \cite{Hoogenboom-2003b, Levy-2008, Piriou-2010}, and
requires to go beyond the formalisms of Refs~\onlinecite{Schrieffer-1963} and
\onlinecite{Pickett-1980}. In particular, one can no longer assume that the
tunneling conductance is proportional to the product of the normal-state DOS by
the ``effective superconducting DOS'' \cite{Schrieffer-1963}
$\text{Re}\big[|\omega|/\sqrt{\omega^2-\Delta^2(\omega)}\big]$, so that nothing
justifies \textit{a priori} to normalize the low-temperature tunneling
conductance by the normal-state conductance as was done with low-$T_c$
superconductors.

Among the new approaches introduced to study strong-coupling effects in HTS,
some have focused on generalizing the classical formalism to the case of
$d$-wave pairing \cite{Abanov-2000, Zasadzinski-2003, Chubukov-2004,
Sandvik-2004}, still overlooking the dimensionality. Other models are strictly
two dimensional (2D) and pay attention to the full electron dispersion
\cite{Eschrig-2000, Hoogenboom-2003b, Devereaux-2004, Zhu-2006, Levy-2008,
Onufrieva-2009}, taking into account the singularities of $N_0(\omega)$. Most of
these studies assume that the collective mode responsible for the
strong-coupling signatures is the sharp $(\pi,\pi)$ spin resonance common to all
cuprates near 30--50 meV (Ref.~\onlinecite{Sidis-2004}), but a phonon scenario
was also put forward \cite{Devereaux-2004, Sandvik-2004}. In the present work,
we extend these approaches to three dimensions (3D) by means of an additional
hopping $t_z$ describing the dispersion along the $c$ axis, and we study the
evolution of the strong-coupling features in the tunneling spectrum along the 2D
to 3D transition on increasing $t_z$. For simplicity we restrict to the
spin-resonance scenario; the electron-phonon model can be treated along the same
lines, and both models lead to the same main conclusions. The model we use is
described in Sec.~\ref{sec:model}, results are presented and discussed in
Sec.~\ref{sec:results}, and Sec.~\ref{sec:NT} is devoted to investigating the
validity and usefulness of the effective superconducting DOS concept.

\section{Model and method}\label{sec:model}

Following previous works \cite{Hoogenboom-2003b, Levy-2008, Jenkins-2009} we
assume that the differential conductance measured by a scanning tunneling
microscope (STM) is proportional to the thermally-broadened local density of
states (LDOS) at the tip apex \cite{Tersoff-1983, Chen-1990b}, and that
furthermore the energy dependence of the LDOS just outside the sample follows
the energy dependence of the bulk DOS $N(\omega)$:
	\begin{equation}\label{eq:dIdV}
		\frac{dI}{dV}\propto\int d\omega\,[-f'(\omega-eV)]N(\omega),
	\end{equation}
where $f'$ is the derivative of the Fermi function. In STM experiments, various
sources of noise may contribute to broaden $N(\omega)$ further; when comparing
theory and experiment we shall take these into account by a phenomenological
Gaussian broadening.

In the superconducting state, the interaction with longitudinal spin
fluctuations is described by the $2\times2$ Nambu matrix self-energy,
	\begin{multline}\label{eq:Sigma}
		\hat{\Sigma}(\vec{k},\omega)=-\frac{1}{\mathscr{N}}\sum_{\vec{q}}
		\frac{1}{\beta}\sum_{i\Omega_n}\,g^2\chi_s(\vec{q},i\Omega_n)\times\\
		\hat{G}_0(\vec{k}-\vec{q},i\omega_n-i\Omega_n)
		\big|_{i\omega_n\to\omega+i0^+}
	\end{multline}
with $\chi_s$ the $\langle S^zS^z\rangle$ spin susceptibility, $g=\sqrt{3}\hbar
J/2$ the coupling strength with $J$ the spin-spin interaction energy, and
$\hat{G}_0$ the $2\times2$ BCS matrix Green's function. The sums extend over the
$\mathscr{N}$ vectors $\vec{q}\equiv(\vec{q}_{\parallel},q_z)$ in the
three-dimensional Brillouin zone and the even Matsubara frequencies
$i\Omega_n=2n\pi/\beta$ with $\beta=(k_{\text{B}}T)^{-1}$.
Equation~(\ref{eq:Sigma}) gives the lowest-order term of an expansion in $J$
\cite{self-consistency}. We follow Ref.~\onlinecite{Eschrig-2000} and use for
$\chi_s$ a model inspired by neutron-scattering experiments on the high-$T_c$
compounds. In this model, $\chi_s$ has no $q_z$ dependence and is the product of
a Lorentzian peak centered at $(\pi,\pi)$ in the 2D Brillouin zone and another
Lorentzian peak centered at the resonance energy $\Omega_{\text{sr}}$. The
widths of the peaks are $\Delta q$ in momentum space and $\Gamma_{\text{sr}}$ in
energy. This simple separable form of $\chi_s$ allows to evaluate analytically
the frequency sum in Eq.~(\ref{eq:Sigma}), and to perform analytically the
continuation from the odd frequencies $i\omega_n=(2n+1)\pi/\beta$ to the
real-frequency axis. The remaining momentum integral is a convolution which can
be efficiently performed using fast Fourier transforms. Also, the absence of
$q_z$ dependence in $\chi_s$ implies that the self-energy does not depend on
$k_z$. We use a BCS Green's function broadened by a small phenomenological
scattering rate $\Gamma$,
	\begin{equation}\label{eq:GBCS}
		\hat{G}_0(\vec{k},i\omega_n)=\frac{[i\omega_n+i\Gamma(i\omega_n)]\hat{\tau}_0+
		\xi_{\vec{k}}\hat{\tau}_3+\Delta_{\vec{k}_{\parallel}}\hat{\tau}_1}
		{[i\omega_n+i\Gamma(i\omega_n)]^2-\xi_{\vec{k}}^2-\Delta_{\vec{k}_{\parallel}}^2},
	\end{equation}
where $\Gamma(i\omega_n)=\Gamma\,\text{sign}(\text{Im}\,i\omega_n)$,
$\hat{\tau}_i$ are the Pauli matrices with $\hat{\tau}_0$ the identity, and
$\Delta_{\vec{k}_{\parallel}}=\Delta_0(\cos k_x-\cos k_y)/2$ is the $d$-wave gap
which we assume $k_z$ independent for simplicity. The additional effects
resulting from a possible weak modulation \cite{Rajagopal-1996, Pairor-2005} of
the gap along $k_z$ will be discussed toward the end of Sec.~\ref{sec:results}.
We do not address here the origin of the pairing leading to the BCS gap
$\Delta_{\vec{k}_{\parallel}}$. With the high-$T_c$ compounds in mind, we
consider a one-band model of quasi-2D electrons with a normal-state dispersion,
	\begin{equation}\label{eq:dispersion}
		\xi_{\vec{k}}=\xi_{\vec{k}_{\parallel}}+2t_z\cos(k_zc),
	\end{equation}
where $\xi_{\vec{k}_{\parallel}}=\varepsilon_{\vec{k}_{\parallel}}-\mu$, $\mu$
being the chemical potential and $\varepsilon_{\vec{k}_{\parallel}}$ a
five-neighbor tight-binding model on the square lattice ($a\equiv1$),
$\varepsilon_{\vec{k}_{\parallel}}=2t_1(\cos k_x+\cos k_y)+4t_2\cos k_x\cos
k_y+2t_3(\cos 2k_x+\cos 2k_y)+4t_4(\cos 2k_x\cos k_y+\cos k_x\cos 2k_y)+4t_5\cos
2k_x\cos 2k_y$.

The momentum dependence of the self-energy in Eq.~(\ref{eq:Sigma}) is not small
(see, e.g., Fig.~1 of Ref.~\onlinecite{Eschrig-2000} and Fig.~\ref{fig:fig2}
below). This is a major difference with respect to the electron-phonon models
describing low-$T_c$ three-dimensional metals, where the momentum dependence of
the self-energy can be neglected. The calculation of the DOS is therefore much
more demanding since two three-dimensional momentum integrations must be
performed for every energy $\omega$. The DOS is given by
	\begin{equation}\label{eq:DOS}
		N(\omega)=\frac{1}{\mathscr{N}}\sum_{\vec{k}}\left({\textstyle-\frac{1}{\pi}}\right)
		\text{Im}\,\hat{G}_{11}(\vec{k},\omega),
	\end{equation}
where $\hat{G}_{11}$ is the first component of the matrix
$\hat{G}(\vec{k},\omega)=[\hat{G}^{-1}_0(\vec{k},\omega)-\hat{\Sigma}(\vec{k},
\omega)]^{-1}$. In Eq.~(\ref{eq:DOS}), the $k_z$ integration can be performed
analytically (see Appendix) but not in Eq.~(\ref{eq:Sigma}). In order to achieve
a good accuracy when computing the DOS $N(\omega)$, we evaluate the self-energy
using a $2048\times2048\times256$ mesh in momentum space and a value
$\Gamma=0.5$~meV. For the evaluation of the tunneling conductance $\Gamma$ is
increased to 2~meV, which allows to decrease the mesh size to
$1024\times1024\times256$.

The model in Eqs.~(\ref{eq:dIdV})--(\ref{eq:DOS}) has several parameters but our
focus is on the $c$-axis hopping energy $t_z$. In HTS, $t_z$ is not larger than
a few meV, and setting it to zero seems appropriate to discuss tunneling data.
Indeed, in the 2D limit the model was found to fit the experimental data for
optimally doped Bi-2223 very well \cite{Levy-2008, Jenkins-2009}. Here we take
the parameters determined from one such fit as a starting point, and we vary
$t_z$ to demonstrate the role of the dimensionality on the tunneling spectrum.
The band parameters $t_{1\ldots5}$ are $-200$, $31$, $-16$, $8$, and $-7$~meV,
and the chemical potential is $\mu=-200$~meV. The gap magnitude is
$\Delta_0=46$~meV. The spin-resonance energy is $\Omega_{\text{sr}}=34$~meV, its
energy width $\Gamma_{\text{sr}}=2$~meV, and its momentum width $\Delta
q=1.6a^{-1}$. Finally the coupling strength is $g=775$~meV, which implies a
quasi-particle residue $Z=0.44$ and a mass renormalization $m^*/m=2.32$ at the
nodal point of the Fermi surface. The temperature is set to $T=2$~K unless
stated otherwise. The resulting theoretical tunneling conductance for $t_z=0$ is
compared with experimental data in Fig.~\ref{fig:fig1}c (topmost curve).

\section{Results}\label{sec:results}

\begin{figure}[t]
\includegraphics[width=\columnwidth]{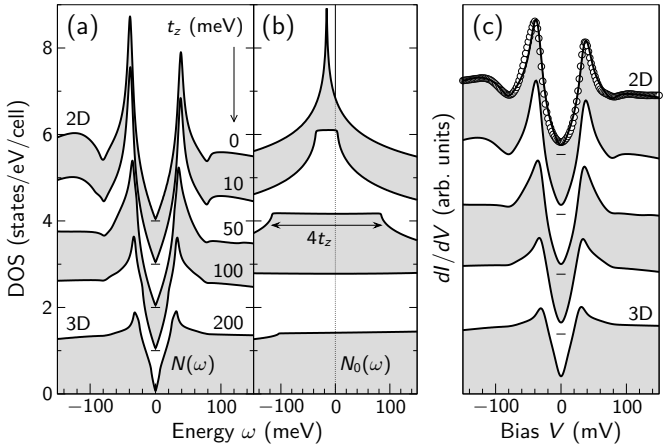}
\caption{\label{fig:fig1}
(a) Density of states $N(\omega)$ for superconducting electrons coupled to a
$(\pi,\pi)$ spin resonance, as a function of the $c$-axis hopping $t_z$. For
$t_z=0$, the system is two dimensional while for $t_z=200$~meV, it is three
dimensional. (b) Normal-state bare DOS $N_0(\omega)$ for the same $t_z$ values.
(c) Tunneling conductance for the same $t_z$ values. The temperature is $T=2$~K
and a Gaussian broadening of 4~meV has been applied. The circles show the
experimental data of Ref.~\onlinecite{Levy-2008} compared to the $t_z=0$
spectrum. The curves in (a), (b), and (c) are offset vertically for clarity.
}
\end{figure}

\begin{figure}[t]
\includegraphics[width=\columnwidth]{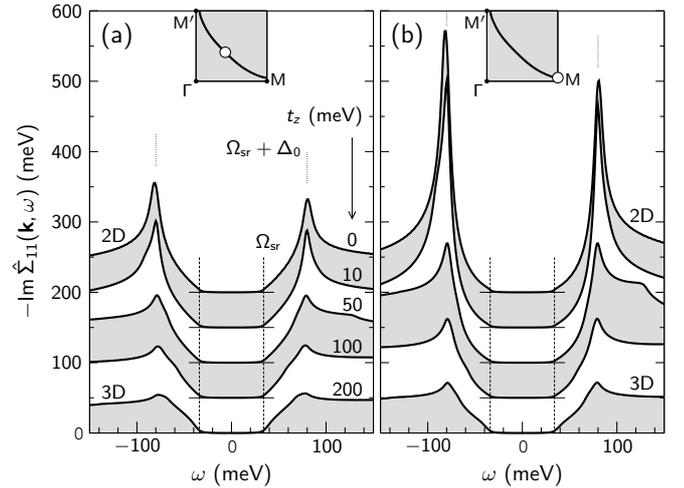}
\caption{\label{fig:fig2}
Scattering rate $-\text{Im}\,\hat{\Sigma}_{11}$ as a function of energy for
several values of $t_z$ at (a) the nodal point
$\vec{k}_{\parallel}\approx(0.41,0.41)\pi/a$ and (b) the anti nodal point
$\vec{k}_{\parallel}\approx(1,0.05)\pi/a$ of the Fermi surface shown in the
insets. Curves are offset vertically for clarity. The dashed vertical lines
delimit the energy range $|\omega|<\Omega_{\text{sr}}$ where inelastic
scattering by the spin resonance is forbidden. The dotted vertical lines
indicate $\pm(\Omega_{\text{sr}}+\Delta_0)$.
}
\end{figure}

The evolution of $N(\omega)$ with increasing $t_z$ is displayed in
Fig.~\ref{fig:fig1}a. In the 2D limit, the DOS shows sharp and particle-hole
asymmetric coherence peaks, strong and asymmetric dips, as well as humps and
shoulders where the spectral weight expelled from the dips is accumulated. This
produces, in particular, a characteristic widening of the coherence peaks basis,
which becomes triangular. The particle-hole asymmetries reflect the
particle-hole asymmetry of the corresponding bare DOS $N_0(\omega)$ shown in
Fig.~\ref{fig:fig1}b, whose Van Hove singularity (VHS) lies slightly below
$E_{\text{F}}$ at $-16$~meV: on the one hand, the spectral weight of the VHS
goes to a larger degree into the negative-energy coherence peak, and on the
other hand the enhancement of the scattering rate due to the VHS is stronger at
negative energy, explaining the stronger dip at $\omega<0$ \cite{Eschrig-2000,
Levy-2008}. This can also be seen in Fig.~\ref{fig:fig2} where the
electron-scattering rate $-\text{Im}\,\hat{\Sigma}_{11}(\vec{k},\omega)$ is
displayed for the nodal and anti nodal points of the Fermi surface. The
scattering rate vanishes for $|\omega|<\Omega_{\text{sr}}$ and has a pronounced,
particle-hole asymmetric maximum near $|\omega|=\Omega_{\text{sr}}+\Delta_0$
(more precisely between $\Omega_{\text{sr}}+\Delta_0$ and
$\Omega_{\text{sr}}+[\xi_{(\pi,0)}^2+\Delta_0^2]^{1/2}$). It is also clear from
the figure that the energy of the scattering-rate peak shows no dispersion with
momentum \cite{Eschrig-2000} but its intensity is strongly momentum dependent
and larger by a factor $\sim 2.5$ in the anti nodal region as compared to the
nodal region.

For $t_z>0$, the logarithmic divergence in $N_0(\omega)$ is cut on the scale of
$4t_z$ due to dispersion along the $c$ axis (Fig.~\ref{fig:fig1}b). No
significant change in either $N(\omega)$ or $dI/dV$ is observed for
$t_z=10$~meV. This value is an upper bound for the $c$-axis hopping energy in
the cuprates, and the relative insensitivity of the DOS to a small $c$-axis
dispersion justifies the use of two-dimensional models for these systems. At
larger $t_z$ values, however, the suppression of the divergence in $N_0(\omega)$
induces a drop of the coherence peaks in $N(\omega)$ and $dI/dV$. This is a
\emph{direct} effect of dimensionality on the tunneling spectrum, which was
overlooked in the conventional strong-coupling approaches of
Refs~\onlinecite{Schrieffer-1963} and \onlinecite{Pickett-1980}.
Simultaneously the peak in the scattering rate is also suppressed with
increasing $t_z$ (Fig.~\ref{fig:fig2}), leading to a weakening of the dip
feature in $N(\omega)$ and $dI/dV$. This is an indirect effect of
dimensionality, that is only revealed in the strong-coupling signatures.

\begin{figure}[b]
\includegraphics[width=\columnwidth]{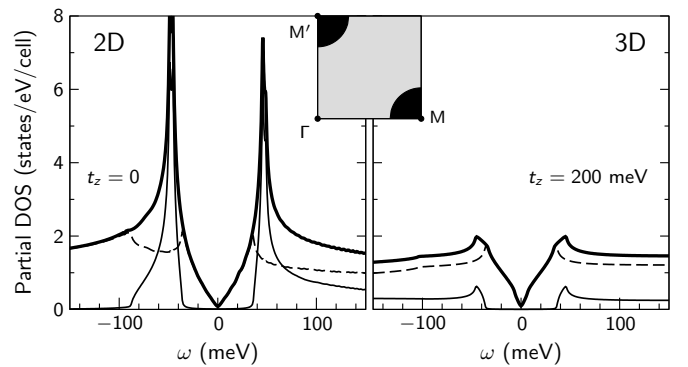}
\caption{\label{fig:fig3}
Partial BCS density of states in two (2D) and three (3D) dimensions. The thin
solid lines show the contribution coming from the $(\pi,0)$ region of the
two-dimensional Brillouin zone, shaded in black in the inset while the dashed
lines show the contribution of the remainder of the zone (shaded in gray). The
thick line is the total BCS DOS.
}
\end{figure}

As Fig.~\ref{fig:fig2} shows, increasing the dimension not only suppresses the
peak at $\Omega_{\text{sr}}+\Delta_0$ in the scattering rate but it also reduces
its momentum dependence. In 2D, this peak arises because the $\vec{q}$ sum in
Eq.~(\ref{eq:Sigma}) is dominated by the saddle-point region near
$\vec{k}_{\text{M}}\equiv(\pi,0)$ and $\vec{k}_{\text{M}'}\equiv(0,\pi)$, where
the spectral weight of the BCS Green's function is largest---\textit{i.e.},
$\vec{k}_{\parallel}-\vec{q}_{\parallel}\approx\vec{k}_{\text{M},\text{M}'}$.
Hence the peak energy is determined chiefly by the BCS excitation energy at
$\vec{k}_{\text{M},\text{M}'}$, shifted by $\Omega_{\text{sr}}$ due to the
convolution with the spin susceptibility, and the peak intensity is controlled
by the momentum dependence of
$\chi_s\big(\vec{k}_{\parallel}-\vec{k}_{\text{M},\text{M}'},\Omega_{\text{sr}}
\big)$, which is at maximum for
$\vec{k}_{\parallel}=\vec{k}_{\text{M}',\text{M}}$. The situation changes in 3D
because the anti-nodal regions no longer dominate the spectral weight, as
illustrated in Fig.~\ref{fig:fig3}. This figure displays the partial BCS density
of states, \textit{i.e.}, the part of the BCS DOS originating from states close
to the $(\pi,0)$ and equivalent points. While in the 2D limit, a region covering
just 14\% of the zone around $(\pi,0)$ provides 56\% of the spectral weight for
energies between $\Omega_{\text{sr}}$ and $\Omega_{\text{sr}}+\Delta_0$, its
contribution is reduced to 21\% in the 3D limit. Hence the scattering rate in 3D
is nearly momentum independent and almost constant above
$\Omega_{\text{sr}}+\Delta_0$. Finally, the 2D to 3D transition also suppresses
the particle-hole asymmetry of the scattering rate. This again results from the
disappearance of particle-hole asymmetry in the underlying bare DOS
(Fig.~\ref{fig:fig1}b) and in the corresponding BCS DOS (Fig.~\ref{fig:fig3}).
Thus the $k_z$ dispersion simultaneously defeats four players who contribute to
make the strong-coupling signatures in the 2D high-$T_c$ superconductors
distinctly different from those in 3D metals: the Van Hove singularity, the
particle-hole asymmetry, the momentum dependence, and the strong scattering
enhancement at $|\omega|\approx\Omega_{\text{sr}}+\Delta_0$, especially near
$(\pi,0)$.

\begin{figure}[t]
\includegraphics[width=\columnwidth]{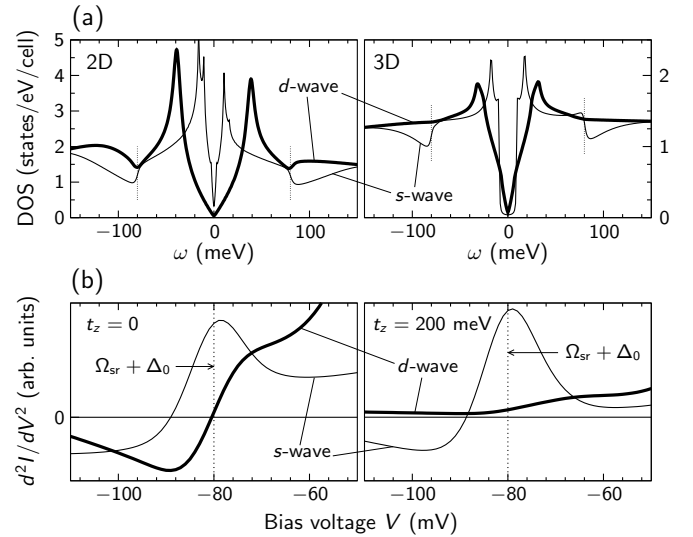}
\caption{\label{fig:fig4}
(a) Comparison of the $d$-wave and $s$-wave DOS for $t_z=0$ (2D) and
$t_z=200$~meV (3D). The thick lines show $N(\omega)$ as in Fig.~\ref{fig:fig1}a.
The thin lines show $N(\omega)$ computed with an $s$-wave gap of the same
magnitude $\Delta_0=46$~meV, and all other parameters unchanged. (b)
Second-derivative tunneling spectrum, $d^2I/dV^2$, in the region of the
strong-coupling signature at negative bias. In the $s$-wave case, there is a
peak in $d^2I/dV^2$ close to the energy $\Omega_{\text{sr}}+\Delta_0$ while in
the $d$-wave case, there is a sign change in 2D and no clear signature in 3D.
}
\end{figure}

In the curves of Figs.~\ref{fig:fig1}a and \ref{fig:fig1}c corresponding to the
3D limit, the strong-coupling signatures are barely visible. Their magnitude is
$\sim1$\%, smaller than the $\sim5$\% value observed in Pb. The origin of this
difference lies in the gap symmetry. In $d$-wave superconductors, the coherence
peaks in the BCS DOS are weak logarithmic singularities \cite{Won-1994} while in
$s$-wave superconductors, they are strong square-root divergences. The strength
of the scattering-rate peak at $\Omega_{\text{sr}}+\Delta_0$, and consequently
the strength of the dip in the DOS and tunneling spectrum, are determined by the
strength of the coherence peaks in the BCS DOS, as is clear from
Eq.~(\ref{eq:Sigma}). In the case of a $d$-wave superconductor, the coherence
peaks are cut in 3D as compared to 2D (see Fig.~\ref{fig:fig3}) in the same way
as the logarithmic VHS in Fig.~\ref{fig:fig1}b, resulting in the suppression of
the scattering-rate enhancement at $\Omega_{\text{sr}}+\Delta_0$ in
Fig.~\ref{fig:fig2}. (Note that, roughly speaking, the scattering rate is
proportional to the BCS DOS shifted in energy by $\pm\Omega_{\text{sr}}$.) The
suppression of the BCS coherence peaks with increasing dimension also occurs in
$s$-wave superconductors, but with one difference: if, on the one hand, the part
of the coherence-peak spectral weight coming from the VHS gets suppressed, on
the other hand, the square-root gap-edge singularities persist in any dimension.
Therefore, in $s$-wave superconductors the strong-coupling signatures remain
clearly visible in 3D. This is illustrated in Fig.~\ref{fig:fig4}a. The 2D and
3D DOS curves of Fig.~\ref{fig:fig1}a are compared to the curves obtained for
the corresponding $s$-wave model, \textit{i.e.}, with all parameters unchanged
except the gap which is replaced by
$\Delta_{\vec{k}_{\parallel}}\equiv\Delta_0=46$~meV. The changes are quite
dramatic. The first effect to notice is a drastic reduction in the peak-to-peak
gap $\Delta_p$ in the $s$-wave case: a consequence of the pair-breaking nature
of the coupling Eq.~(\ref{eq:Sigma}) in the $s$-wave channel \cite{Annett-1990,
endnote1}. Still, the strong-coupling signatures appear at the same energy
$\Omega_{\text{sr}}+\Delta_0=80$~meV in both $d$ and $s$ wave, due to our choice
of the lowest-order model $\hat{\Sigma}\propto\chi_s\hat{G}_0$ in
Eq.~(\ref{eq:Sigma}). The second observation is that the strong-coupling
signatures look like steps in the $s$-wave DOS, like in the classical
superconductors \cite{Schrieffer-1963}, reflecting the asymmetric shape of the
BCS $s$-wave coherence peaks. In contrast, the signatures appear as local minima
in the $d$-wave DOS, because the coherence peaks of the $d$-wave BCS DOS are
nearly symmetric about their maximum. In short, the strong-coupling features
give an ``inverted image'' of the BCS coherence peaks \cite{Levy-2008}. An
interesting consequence follows: while in $s$-wave superconductors, the
strong-coupling structures correspond to \emph{peaks} in the second-derivative
$d^2I/dV^2$ spectrum, for a $d$-wave gap they correspond to \emph{zeros} in the
$d^2I/dV^2$ spectrum, as demonstrated in Fig.~\ref{fig:fig4}b. This conclusion
applies equally to phonon models and calls for a reinterpretation of cuprate
$d^2I/dV^2$ data in which peaks were assigned to phonon modes \cite{Lee-2006,
Zhao-2007}. Finally, one sees from Fig.~\ref{fig:fig4} that in 3D the signatures
remain strong for an $s$-wave gap, for the reason explained above, while they
have almost disappeared in the $d$-wave case.

\begin{figure}[t]
\includegraphics[width=\columnwidth]{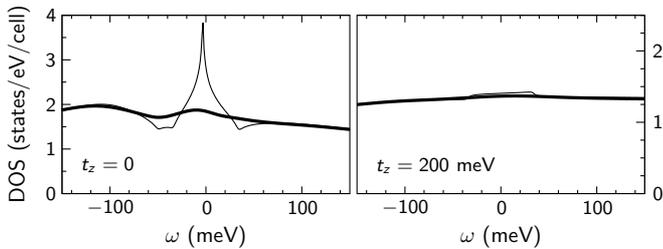}
\caption{\label{fig:fig5}
Normal-state DOS. The thin lines show the $T=2$~K DOS for $\Delta_0=0$; the
thick lines show the $T=200$~K DOS for $\Delta_0=0$ and
$\Gamma_{\text{sr}}=14$~meV (see text). All other parameters are as in
Fig.~\ref{fig:fig1}a.
}
\end{figure}

The previous discussion underlines the role of the BCS coherence peaks in the
formation, strength, and shape of the strong-coupling signatures. More
generally, for such signatures to occur there must be divergences (or at least
pronounced maxima) in the non-interacting DOS. Peaks in the ``bosonic'' spectrum
are not sufficient, although they are necessary. Indeed, phonon structures are
absent from the normal-state spectra of classical superconductors
\cite{McMillan-1969} because the normal-state DOS is flat, in spite of the facts
that the phonon spectrum and the electron-phonon coupling do not change
significantly at $T_c$. In contrast, the normal-state DOS of 2D high-$T_c$
superconductors exhibits structures, either the pseudogap \cite{Fischer-2007} or
the bare VHS \cite{Piriou-2010}. One can therefore expect to see strong-coupling
features in the normal-state spectra of HTS, provided that the peaks in the
bosonic spectrum subsist above $T_c$. Figure~\ref{fig:fig5} (thin lines) shows
the normal-state DOS implied by setting $\Delta_0=0$ in our model, keeping the
other parameters fixed (including temperature). As expected sharp
strong-coupling features remain in 2D at energies $\pm\Omega_{\text{sr}}$ and
$\xi_{(\pi,0)}-\Omega_{\text{sr}}=-50$~meV while nothing but very weak
structures subsist in 3D, signaling the onset of scattering at
$\pm\Omega_{\text{sr}}$. Unfortunately it turns out that in the HTS the spin
resonance is absent above $T_c$---or at least below the background level of
neutron-scattering experiments \cite{Bourges-2000a}. The normal state of Bi-2223
has not been investigated by neutron scattering so far but we may borrow
information from the much studied YBa$_2$Cu$_3$O$_{6+x}$ system (Y-123). In
Y-123, the normal-state spin susceptibility preserves its separable form with
independent momentum and energy variations \cite{Regnault-1995}. It is still
centered at $(\pi,\pi)$ with a broad maximum at a characteristic
temperature-dependent frequency $\Omega_{\text{sf}}\approx\Omega_{\text{sr}}$.
For the purpose of illustrating the effect of a broad spin-fluctuations
continuum on the normal-state tunneling spectrum, it is sufficient to use the
same model as in the superconducting state but with the new parameter
$\Gamma_{\text{sr}}=14$~meV \cite{endnote2}. The resulting DOS calculated at
$T=200$~K is shown by the thick lines in Fig.~\ref{fig:fig5}. The
strong-coupling signatures are almost washed out in 2D and completely in 3D.
This is not due to the thermal broadening of Eq.~(\ref{eq:dIdV}), not included
in the DOS $N(\omega)$, but mostly to the \emph{intrinsic} temperature
dependence of the self-energy in Eq.~(\ref{eq:Sigma}), and, to a lesser extent,
to the broader spin response. Hence, if structures due to interaction with spin
fluctuations are unlikely to show up in the normal state of HTS, those
associated with the interaction with phonons may well be observable if the
coupling is strong enough since this coupling will not change appreciably at
$T_c$.

In the present study, we have overlooked a possible $k_z$ dependence of the BCS
gap, retaining only the $k_z$ dependence of the bare dispersion. A weak
modulation of the BCS gap along $k_z$ is expected in 3D systems
\cite{Rajagopal-1996}. As shown in Ref.~\onlinecite{Pairor-2005}, such a
modulation has the effect of cutting the logarithmic coherence peaks on the
scale of $2\Delta_z$, with $\Delta_z$ the amplitude of the gap modulation. This
is similar to the effect of $t_z$ on the BCS coherence peaks, which are cut on a
scale corresponding to the gap variation along the warped 3D Fermi surface,
namely, $\sim\Delta_0t_z/4t_1$, as seen in Fig.~\ref{fig:fig3}. The expected
effect of $\Delta_z$ on the scattering rate is also an additional broadening on
top of the one produced by $t_z$, $\hat{G}_0$ being replaced by its $k_z$
average in Eq.~(\ref{eq:Sigma}). Therefore, we expect that the gap modulation
along $k_z$ will contribute to suppress the coherence peaks and the
strong-coupling features even further with increasing $t_z$, as compared to the
results in Fig.~\ref{fig:fig1}.

Our results can be summarized as follows. The formation of clear strong-coupling
structures in the tunneling conductance requires two ingredients: (A) at least
one peak in the spectrum of collective excitations and (B) at least one peak in
the non-interacting or superconducting DOS. In classical superconductors, (A) is
provided by optical phonons and (B) is the asymmetric square-root singularity at
the edge of the $s$-wave gap: strong-coupling features are asymmetric
steps---peaks in the $d^2I/dV^2$ curve---and dimensionality plays no big role
because (A) and (B) are present in any dimension. In the normal state, there is
no signature because (B) is absent. In high-$T_c$ layered superconductors, (A)
is provided by the spin resonance and (B) has two sources: (B1) the logarithmic
Van Hove singularity in the bare DOS; (B2) the symmetric logarithmic
singularities at the edge of the $d$-wave gap. Strong-coupling signatures appear
as local minima---zeros in the $d^2I/dV^2$ curve---but they vanish with
increasing dimensionality from 2D to 3D because (B1) and (B2) both get
suppressed by the $c$-axis dispersion. In the normal state of two-dimensional
HTS, (B2) is absent, leaving aside the question of the pseudogap but (B1)
remains and strong-coupling signatures are thus expected unless (A) disappears
at $T_c$. This is the case for the spin resonance but certainly not for phonons,
leaving open the possibility that phonon structures might be observable in the
normal-state tunneling spectra.

\section{DOS and effective DOS}\label{sec:NT}

The conventional theory of electron tunneling into superconductors
\cite{Schrieffer-1963} leads to an equation identical to Eq.~(\ref{eq:dIdV}) for
the tunneling conductance, except that the DOS $N(\omega)$ is replaced by an
``effective tunneling DOS''
$N_T(\omega)=N_0(0)\text{Re}\big[|\omega|/\sqrt{\omega^2-\Delta^2(\omega)}\big]$.
$N_0(0)$ is the normal-state DOS at zero energy---$N_0(\omega)\equiv N_0(0)$
is assumed---and $\Delta(\omega)$ is the gap function. The latter must be
understood as a Fermi-surface average of weakly momentum-dependent quantities,
$\Delta(\omega)=\langle\Phi(\vec{k},\omega)/Z(\vec{k},\omega)\rangle_{\text{FS}}
$ with $\Phi$ and $Z$ the Eliashberg pairing and renormalization functions. In
the notation of Eq.~(\ref{eq:Sigma}), they read
$Z(\vec{k},\omega)=1-[\hat{\Sigma}_{11}(\vec{k},\omega)+\hat{\Sigma}_{22}(\vec{k
},\omega)]/(2\omega)$ and, for an $s$-wave gap $\Delta_0$,
$\Phi(\vec{k},\omega)=\Delta_0+\hat{\Sigma}_{12}(\vec{k},\omega)$. In a $d$-wave
superconductor, the Fermi-surface average of the gap $\Delta_{\vec{k}}$
vanishes, and so does the average of the off-diagonal self-energy since
$\hat{\Sigma}_{12}(\vec{k},\omega)\propto\Delta_{\vec{k}}$. The effective
tunneling DOS concept is logically generalized \cite{Abanov-2000,
Zasadzinski-2003} by writing
$N_T(\omega)=N_0(0)\text{Re}\big\langle|\omega|/\sqrt{\omega^2-[\Delta_{\vec{k}}
\phi(\omega)]^2}\big\rangle_{\text{FS}}$ with
	\begin{equation}\label{eq:Psi}
		\phi(\omega)=\left\langle
		\frac{1+\hat{\Sigma}_{12}(\vec{k},\omega)/\Delta_{\vec{k}}}
		{1-[\hat{\Sigma}_{11}(\vec{k},\omega)+\hat{\Sigma}_{22}(\vec{k},\omega)]/(2\omega)}
		\right\rangle_{\text{FS}}.
	\end{equation}
This form of $N_T(\omega)$ is an even function of $\omega$, and cannot fit the
particle-hole asymmetric spectra in HTS. Therefore, a further generalization of
the effective tunneling DOS has been necessary, namely,
	\begin{equation}\label{eq:NT}
		N_T(\omega)=N_0(\omega)\,\text{Re}\left\langle\frac{|\omega|}
		{\sqrt{\omega^2-[\Delta_{\vec{k}}\phi(\omega)]^2}}\right\rangle_{\text{FS}}
	\end{equation}
which suggests that the ``true'' superconducting DOS can be obtained by dividing
the tunneling spectrum in the superconducting state by the spectrum in the
normal state \cite{Zasadzinski-2003, Romano-2006, Boyer-2007, Pasupathy-2008}.

Equation~(\ref{eq:NT}) is very convenient, but lacks a formal justification. Our
model offers the opportunity to investigate the usefulness of Eq.~(\ref{eq:NT}),
by comparing numerically the actual tunneling DOS $N(\omega)$ of
Eq.~(\ref{eq:DOS}) with the effective tunneling DOS $N_T(\omega)$. For the
practical evaluation of $N_T(\omega)$, we define the Fermi-surface average as
	\begin{equation}
		\left\langle\,\cdots\right\rangle_{\text{FS}}\equiv\frac{\sum_{\vec{k}}
		A_0(\vec{k},0)\,(\,\cdots)}{\sum_{\vec{k}}A_0(\vec{k},0)}
	\end{equation}
with $A_0(\vec{k},0)$ the zero-energy spectral function in the absence of
pairing:
$A_0(\vec{k},0)=(-1/\pi)\text{Im}\,\hat{G}_{11}(\vec{k},0)|_{\Delta_{\vec{k}}=0}
$. With this definition, the average is performed on the \emph{renormalized}
Fermi surface, defined by
$\xi_{\vec{k}}+\text{Re}\,\hat{\Sigma}_{11}(\vec{k},0)=0$, rather than the bare
Fermi surface $\xi_{\vec{k}}=0$. Furthermore, each state gets correctly weighted
if the spectral weight is unevenly distributed along the Fermi surface.

\begin{figure}[t]
\includegraphics[width=\columnwidth]{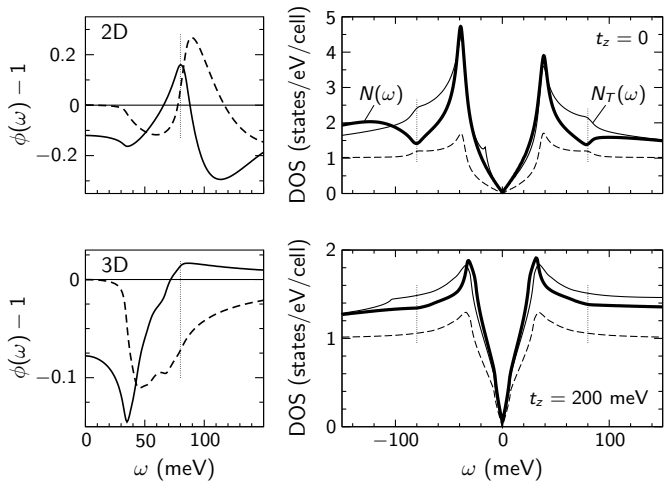}
\caption{\label{fig:fig6}
(Left panels) Real part (solid lines) and imaginary part (dashed lines) of the
pairing function defined in Eq.~(\ref{eq:Psi}) for $t_z=0$ (2D) and
$t_z=200$~meV (3D). (Right panels) Comparison of the effective tunneling DOS
$N_T(\omega)$ of Eq.~(\ref{eq:NT}) with the actual DOS $N(\omega)$ of
Eq.~(\ref{eq:DOS}). The dashed lines show $N_T(\omega)/N_0(\omega)$. In all
graphs, the dotted vertical lines mark the energy
$\pm(\Omega_{\text{sr}}+\Delta_0)$.
}
\end{figure}

A comparison of $N(\omega)$ and $N_T(\omega)$ is displayed in
Fig.~\ref{fig:fig6}, where the pairing function $\phi(\omega)$ is also shown. In
two dimensions, the real part of $\phi(\omega)$ has a maximum at
$\omega=\Omega_{\text{sr}}+\Delta_0$, where its imaginary part shows a rapid
variation. This is analogous to the behavior reported in
Ref.~\onlinecite{Schrieffer-1963}. The resulting $N_T(\omega)$ also shows a
behavior similar to the one found in Ref.~\onlinecite{Schrieffer-1963}:
$N_T(\omega)$ is \emph{larger} than the BCS density of states at energies
smaller than $\Omega_{\text{sr}}+\Delta_0$ and drops below the BCS DOS at
$\Omega_{\text{sr}}+\Delta_0$. The actual DOS $N(\omega)$, however, behaves
differently: it is \emph{smaller} than the BCS DOS between the coherence peak
and some energy above the dip minimum (see also Fig.~1 of
Ref.~\onlinecite{Levy-2008}). Thus, although the positions of the
strong-coupling features are identical in $N_T(\omega)$ and $N(\omega)$, their
shape is markedly different in 2D $d$-wave superconductors. In 3D, the
difference between $N_T(\omega)$ and $N(\omega)$ is less severe than in 2D, and
both curves show very weak signatures, although those in $N(\omega)$ are
slightly stronger. Finally, the $N_T(\omega)$ curves show structures which are
absent in the $N(\omega)$ curves. In 2D, a peak at
$\omega=-16~\text{meV}=\xi_{(\pi,0)}$ appears due to the VHS in $N_0(\omega)$;
this peak is unphysical because in the actual energy spectrum, the VHS is pushed
to $-[\xi_{(\pi,0)}^2+\Delta_0^2]^{1/2}$. In 3D, $N_T(\omega)$ has a structure
near $-100$~meV, which also comes from the bare DOS $N_0(\omega)$ as can be seen
in Fig.~\ref{fig:fig1}b. In the actual spectrum, this structure is suppressed
due to the persistence of a large scattering rate at energies much higher than
the threshold $\Omega_{\text{sr}}$ (see Fig.~\ref{fig:fig2}). These problems
illustrate the limitations of the simple product Ansatz Eq.~(\ref{eq:NT}) for
analyzing the tunneling spectrum of $d$-wave superconductors.

\acknowledgments

I thank John Zasadzinski for useful discussions. This work was supported by the
Swiss National Science Foundation through Division II and MaNEP.

\appendix

\section*{Appendix: Analytical $k_z$ integration}

If the Nambu self-energy has no $k_z$ dependence, the $k_z$ sum in
Eq.~(\ref{eq:DOS}) can be performed analytically. This is the case in our model
defined in Eq.~(\ref{eq:Sigma}). Solving Dyson's equation
$\hat{G}(\vec{k},\omega)=[\hat{G}^{-1}_0(\vec{k},\omega)-\hat{\Sigma}(\vec{k},
\omega)]^{-1}$ with $\hat{G}_0$ given by Eq.~(\ref{eq:GBCS}), we find
\begin{widetext}
	\begin{equation}
		\hat{G}_{11}(\vec{k},\omega)=	\frac{1}{
		\omega+i\Gamma-\xi_{\vec{k}}-\hat{\Sigma}_{11}(\vec{k},\omega)-
		[\Delta_{\vec{k}_{\parallel}}+\hat{\Sigma}_{12}(\vec{k},\omega)]^2
		\big/[\omega+i\Gamma+\xi_{\vec{k}}-\hat{\Sigma}_{22}(\vec{k},\omega)]}.
	\end{equation}
Since $\hat{\Sigma}$ does not depend on $k_z$ (although it \emph{does} depend on
$t_z$), the $k_z$ dependence only comes from $\xi_{\vec{k}}$ in
Eq.~(\ref{eq:dispersion}) and we can make it explicit by rewriting
	\begin{multline}\label{eq:G11}
		\hat{G}_{11}(\vec{k},\omega)=\frac{1}{z_{11}-2t_z\cos(k_zc)
		-z_{12}^2\big/[z_{22}+2t_z\cos(k_zc)]}\\
		=\frac{1}{2}\left(1+\frac{\zeta}{\lambda}\right)\frac{1}{\eta+\lambda-2t_z\cos(k_zc)}+
		\frac{1}{2}\left(1-\frac{\zeta}{\lambda}\right)\frac{1}{\eta-\lambda-2t_z\cos(k_zc)}.
	\end{multline}
In Eq.~(\ref{eq:G11}), the quantities $z_{11}$, $z_{22}$, $z_{12}$, $\zeta$,
$\lambda$, and $\eta$ are all functions of $\vec{k}_{\parallel}$ and $\omega$
but not of $k_z$. Explicitly,
$z_{11}=\omega+i\Gamma-\xi_{\vec{k}_{\parallel}}-\hat{\Sigma}_{11}(\vec{k}_{
\parallel},\omega)$,
$z_{22}=\omega+i\Gamma+\xi_{\vec{k}_{\parallel}}-\hat{\Sigma}_{22}(\vec{k}_{
\parallel},\omega)$,
$z_{12}=\Delta_{\vec{k}_{\parallel}}+\hat{\Sigma}_{12}(\vec{k}_{\parallel},
\omega)$, $\zeta=(z_{11}+z_{22})/2$, $\lambda=\sqrt{\zeta^2-z_{12}^2}$, and
$\eta=(z_{11}-z_{22})/2$.
\end{widetext}
The $k_z$ integration can then be performed by means of the identity, \\[-1.5em]
	\begin{equation}
		\tilde{D}_{t}(z)\equiv\frac{1}{2\pi}\int_{-\pi}^{\pi}\frac{dx}{z-2t\cos x}
		=\frac{1}{\sqrt{z-2t}\sqrt{z+2t}}
	\end{equation}
and yields
	\begin{multline}
		N(\omega)=\frac{1}{\mathscr{N}_{\parallel}}\sum_{\vec{k}_{\parallel}}
		\left({\textstyle-\frac{1}{\pi}}\right)\text{Im}\,\left\{
		\frac{1}{2}\left(1+\frac{\zeta}{\lambda}\right)\tilde{D}_{t_z}(\eta+\lambda)+\right.\\
		\left.\frac{1}{2}\left(1-\frac{\zeta}{\lambda}\right)\tilde{D}_{t_z}(\eta-\lambda)\right\},
	\end{multline}
where $\mathscr{N}_{\parallel}$ is the number of $\vec{k}_{\parallel}$ points in
the 2D Brillouin zone.


\begin{thebibliography}{47}
\expandafter\ifx\csname natexlab\endcsname\relax\def\natexlab#1{#1}\fi
\expandafter\ifx\csname bibnamefont\endcsname\relax
  \def\bibnamefont#1{#1}\fi
\expandafter\ifx\csname bibfnamefont\endcsname\relax
  \def\bibfnamefont#1{#1}\fi
\expandafter\ifx\csname citenamefont\endcsname\relax
  \def\citenamefont#1{#1}\fi
\expandafter\ifx\csname url\endcsname\relax
  \def\url#1{\texttt{#1}}\fi
\expandafter\ifx\csname urlprefix\endcsname\relax\def\urlprefix{URL }\fi
\providecommand{\bibinfo}[2]{#2}
\providecommand{\eprint}[2][]{\url{#2}}

\bibitem[{\citenamefont{Damascelli et~al.}(2003)\citenamefont{Damascelli,
  Hussain, and Shen}}]{Damascelli-2003}
\bibinfo{author}{\bibfnamefont{A.}~\bibnamefont{Damascelli}},
  \bibinfo{author}{\bibfnamefont{Z.}~\bibnamefont{Hussain}}, \bibnamefont{and}
  \bibinfo{author}{\bibfnamefont{Z.-X.} \bibnamefont{Shen}},
  \bibinfo{journal}{Rev. Mod. Phys.} \textbf{\bibinfo{volume}{75}},
  \bibinfo{pages}{473} (\bibinfo{year}{2003}).

\bibitem[{\citenamefont{Campuzano et~al.}(2004)\citenamefont{Campuzano, Norman,
  and Randeria}}]{Campuzano-2004}
\bibinfo{author}{\bibfnamefont{J.~C.} \bibnamefont{Campuzano}},
  \bibinfo{author}{\bibfnamefont{M.}~\bibnamefont{Norman}}, \bibnamefont{and}
  \bibinfo{author}{\bibfnamefont{M.}~\bibnamefont{Randeria}}, in
  \emph{\bibinfo{booktitle}{Physics of Superconductors}}, edited by
  \bibinfo{editor}{\bibfnamefont{K.~H.} \bibnamefont{Bennemann}}
  \bibnamefont{and} \bibinfo{editor}{\bibfnamefont{J.~B.}
  \bibnamefont{Ketterson}} (\bibinfo{publisher}{Springer},
  \bibinfo{address}{Berlin}, \bibinfo{year}{2004}), Vol.~\bibinfo{volume}{II},
  p. \bibinfo{pages}{167}.

\bibitem[{\citenamefont{Fischer et~al.}(2007)\citenamefont{Fischer, Kugler,
  Maggio-Aprile, Berthod, and Renner}}]{Fischer-2007}
\bibinfo{author}{\bibfnamefont{{\O}.}~\bibnamefont{Fischer}},
  \bibinfo{author}{\bibfnamefont{M.}~\bibnamefont{Kugler}},
  \bibinfo{author}{\bibfnamefont{I.}~\bibnamefont{Maggio-Aprile}},
  \bibinfo{author}{\bibfnamefont{C.}~\bibnamefont{Berthod}}, \bibnamefont{and}
  \bibinfo{author}{\bibfnamefont{C.}~\bibnamefont{Renner}},
  \bibinfo{journal}{Rev. Mod. Phys.} \textbf{\bibinfo{volume}{79}},
  \bibinfo{pages}{353} (\bibinfo{year}{2007}).

\bibitem[{\citenamefont{Eliashberg}(1960)}]{Eliashberg-1960}
\bibinfo{author}{\bibfnamefont{G.~M.} \bibnamefont{Eliashberg}},
  \bibinfo{journal}{Sov. Phys. JETP} \textbf{\bibinfo{volume}{11}},
  \bibinfo{pages}{696} (\bibinfo{year}{1960}).

\bibitem[{\citenamefont{Schrieffer}(1964)}]{Schrieffer-1964b}
\bibinfo{author}{\bibfnamefont{J.~R.} \bibnamefont{Schrieffer}},
  \emph{\bibinfo{title}{Theory of Superconductivity}}
  (\bibinfo{publisher}{Benjamin}, \bibinfo{address}{New York},
  \bibinfo{year}{1964}).

\bibitem[{\citenamefont{Carbotte}(1990)}]{Carbotte-1990}
\bibinfo{author}{\bibfnamefont{J.~P.} \bibnamefont{Carbotte}},
  \bibinfo{journal}{Rev. Mod. Phys.} \textbf{\bibinfo{volume}{62}},
  \bibinfo{pages}{1027} (\bibinfo{year}{1990}).

\bibitem[{\citenamefont{Eschrig and Norman}(2000)}]{Eschrig-2000}
\bibinfo{author}{\bibfnamefont{M.}~\bibnamefont{Eschrig}} \bibnamefont{and}
  \bibinfo{author}{\bibfnamefont{M.~R.} \bibnamefont{Norman}},
  \bibinfo{journal}{Phys. Rev. Lett.} \textbf{\bibinfo{volume}{85}},
  \bibinfo{pages}{3261} (\bibinfo{year}{2000});
  \bibinfo{journal}{Phys. Rev. B} \textbf{\bibinfo{volume}{67}},
  \bibinfo{pages}{144503} (\bibinfo{year}{2003}).

\bibitem[{\citenamefont{Levy~de Castro et~al.}(2008)\citenamefont{Levy~de
  Castro, Berthod, Piriou, Giannini, and Fischer}}]{Levy-2008}
\bibinfo{author}{\bibfnamefont{G.}~\bibnamefont{Levy~de Castro}},
  \bibinfo{author}{\bibfnamefont{C.}~\bibnamefont{Berthod}},
  \bibinfo{author}{\bibfnamefont{A.}~\bibnamefont{Piriou}},
  \bibinfo{author}{\bibfnamefont{E.}~\bibnamefont{Giannini}}, \bibnamefont{and}
  \bibinfo{author}{\bibfnamefont{{\O}.}~\bibnamefont{Fischer}},
  \bibinfo{journal}{Phys. Rev. Lett.} \textbf{\bibinfo{volume}{101}},
  \bibinfo{pages}{267004} (\bibinfo{year}{2008}).

\bibitem[{\citenamefont{Kugler et~al.}(2006)\citenamefont{Kugler, Levy~de
  Castro, Giannini, Piriou, Manuel, Hess, and Fischer}}]{Kugler-2006}
\bibinfo{author}{\bibfnamefont{M.}~\bibnamefont{Kugler}},
  \bibinfo{author}{\bibfnamefont{G.}~\bibnamefont{Levy~de Castro}},
  \bibinfo{author}{\bibfnamefont{E.}~\bibnamefont{Giannini}},
  \bibinfo{author}{\bibfnamefont{A.}~\bibnamefont{Piriou}},
  \bibinfo{author}{\bibfnamefont{A.~A.} \bibnamefont{Manuel}},
  \bibinfo{author}{\bibfnamefont{C.}~\bibnamefont{Hess}}, \bibnamefont{and}
  \bibinfo{author}{\bibfnamefont{{\O}.}~\bibnamefont{Fischer}},
  \bibinfo{journal}{J. Phys. Chem. Solids} \textbf{\bibinfo{volume}{67}},
  \bibinfo{pages}{353} (\bibinfo{year}{2006}).

\bibitem[{\citenamefont{Swihart et~al.}(1965)\citenamefont{Swihart, Scalapino,
  and Wada}}]{Swihart-1965}
\bibinfo{author}{\bibfnamefont{J.~C.} \bibnamefont{Swihart}},
  \bibinfo{author}{\bibfnamefont{D.~J.} \bibnamefont{Scalapino}},
  \bibnamefont{and} \bibinfo{author}{\bibfnamefont{Y.}~\bibnamefont{Wada}},
  \bibinfo{journal}{Phys. Rev. Lett.} \textbf{\bibinfo{volume}{14}},
  \bibinfo{pages}{106} (\bibinfo{year}{1965}).

\bibitem[{\citenamefont{Johnson et~al.}(2001)\citenamefont{Johnson, Valla,
  Fedorov, Yusof, Wells, Li, Moodenbaugh, Gu, Koshizuka, Kendziora
  et~al.}}]{Johnson-2001}
\bibinfo{author}{\bibfnamefont{P.~D.} \bibnamefont{Johnson}},
  \bibinfo{author}{\bibfnamefont{T.}~\bibnamefont{Valla}},
  \bibinfo{author}{\bibfnamefont{A.~V.} \bibnamefont{Fedorov}},
  \bibinfo{author}{\bibfnamefont{Z.}~\bibnamefont{Yusof}},
  \bibinfo{author}{\bibfnamefont{B.~O.} \bibnamefont{Wells}},
  \bibinfo{author}{\bibfnamefont{Q.}~\bibnamefont{Li}},
  \bibinfo{author}{\bibfnamefont{A.~R.} \bibnamefont{Moodenbaugh}},
  \bibinfo{author}{\bibfnamefont{G.~D.} \bibnamefont{Gu}},
  \bibinfo{author}{\bibfnamefont{N.}~\bibnamefont{Koshizuka}},
  \bibinfo{author}{\bibfnamefont{C.}~\bibnamefont{Kendziora}},
  \bibinfo{author}{\bibfnamefont{S.}~\bibnamefont{Jian}}, \bibnamefont{and}
  \bibinfo{author}{\bibfnamefont{D. G.}~\bibnamefont{Hinks}},
  \bibinfo{journal}{Phys. Rev. Lett.}
  \textbf{\bibinfo{volume}{87}}, \bibinfo{pages}{177007}
  (\bibinfo{year}{2001}).

\bibitem[{\citenamefont{Hwang et~al.}(2007)\citenamefont{Hwang, Timusk,
  Schachinger, and Carbotte}}]{Hwang-2007}
\bibinfo{author}{\bibfnamefont{J.}~\bibnamefont{Hwang}},
  \bibinfo{author}{\bibfnamefont{T.}~\bibnamefont{Timusk}},
  \bibinfo{author}{\bibfnamefont{E.}~\bibnamefont{Schachinger}},
  \bibnamefont{and} \bibinfo{author}{\bibfnamefont{J.~P.}
  \bibnamefont{Carbotte}}, \bibinfo{journal}{Phys. Rev. B}
  \textbf{\bibinfo{volume}{75}}, \bibinfo{pages}{144508}
  (\bibinfo{year}{2007}).

\bibitem[{\citenamefont{Schrieffer et~al.}(1963)\citenamefont{Schrieffer,
  Scalapino, and Wilkins}}]{Schrieffer-1963}
\bibinfo{author}{\bibfnamefont{J.~R.} \bibnamefont{Schrieffer}},
  \bibinfo{author}{\bibfnamefont{D.~J.} \bibnamefont{Scalapino}},
  \bibnamefont{and} \bibinfo{author}{\bibfnamefont{J.~W.}
  \bibnamefont{Wilkins}}, \bibinfo{journal}{Phys. Rev. Lett.}
  \textbf{\bibinfo{volume}{10}}, \bibinfo{pages}{336} (\bibinfo{year}{1963}).

\bibitem[{\citenamefont{Scalapino and Anderson}(1964)}]{Scalapino-1964}
\bibinfo{author}{\bibfnamefont{D.~J.} \bibnamefont{Scalapino}}
  \bibnamefont{and} \bibinfo{author}{\bibfnamefont{P.~W.}
  \bibnamefont{Anderson}}, \bibinfo{journal}{Phys. Rev.}
  \textbf{\bibinfo{volume}{133}}, \bibinfo{pages}{A921} (\bibinfo{year}{1964}).

\bibitem[{\citenamefont{McMillan and Rowell}(1965)}]{McMillan-1965}
\bibinfo{author}{\bibfnamefont{W.~L.} \bibnamefont{McMillan}} \bibnamefont{and}
  \bibinfo{author}{\bibfnamefont{J.~M.} \bibnamefont{Rowell}},
  \bibinfo{journal}{Phys. Rev. Lett.} \textbf{\bibinfo{volume}{14}},
  \bibinfo{pages}{108} (\bibinfo{year}{1965}).

\bibitem[{\citenamefont{Pickett}(1980)}]{Pickett-1980}
\bibinfo{author}{\bibfnamefont{W.~E.} \bibnamefont{Pickett}},
  \bibinfo{journal}{Phys. Rev. B} \textbf{\bibinfo{volume}{21}},
  \bibinfo{pages}{3897} (\bibinfo{year}{1980}).

\bibitem[{\citenamefont{Hoogenboom et~al.}(2003)\citenamefont{Hoogenboom,
  Berthod, Peter, Fischer, and Kordyuk}}]{Hoogenboom-2003b}
\bibinfo{author}{\bibfnamefont{B.~W.} \bibnamefont{Hoogenboom}},
  \bibinfo{author}{\bibfnamefont{C.}~\bibnamefont{Berthod}},
  \bibinfo{author}{\bibfnamefont{M.}~\bibnamefont{Peter}},
  \bibinfo{author}{\bibfnamefont{{\O}.}~\bibnamefont{Fischer}},
  \bibnamefont{and} \bibinfo{author}{\bibfnamefont{A.~A.}
  \bibnamefont{Kordyuk}}, \bibinfo{journal}{Phys. Rev. B}
  \textbf{\bibinfo{volume}{67}}, \bibinfo{pages}{224502}
  (\bibinfo{year}{2003}).

\bibitem[{Pir()}]{Piriou-2010}
\bibinfo{note}{A. Piriou, N. Jenkins, C. Berthod, I. Maggio-Aprile, and {\O}.
Fischer, to be published.}

\bibitem[{\citenamefont{Abanov and Chubukov}(2000)}]{Abanov-2000}
\bibinfo{author}{\bibfnamefont{A.}~\bibnamefont{Abanov}} \bibnamefont{and}
  \bibinfo{author}{\bibfnamefont{A.~V.} \bibnamefont{Chubukov}},
  \bibinfo{journal}{Phys. Rev. B} \textbf{\bibinfo{volume}{61}},
  \bibinfo{pages}{R9241} (\bibinfo{year}{2000}).

\bibitem[{\citenamefont{Zasadzinski et~al.}(2003)\citenamefont{Zasadzinski,
  Coffey, Romano, and Yusof}}]{Zasadzinski-2003}
\bibinfo{author}{\bibfnamefont{J.~F.} \bibnamefont{Zasadzinski}},
  \bibinfo{author}{\bibfnamefont{L.}~\bibnamefont{Coffey}},
  \bibinfo{author}{\bibfnamefont{P.}~\bibnamefont{Romano}}, \bibnamefont{and}
  \bibinfo{author}{\bibfnamefont{Z.}~\bibnamefont{Yusof}},
  \bibinfo{journal}{Phys. Rev. B} \textbf{\bibinfo{volume}{68}},
  \bibinfo{pages}{180504(R)} (\bibinfo{year}{2003}).

\bibitem[{\citenamefont{Chubukov and Norman}(2004)}]{Chubukov-2004}
\bibinfo{author}{\bibfnamefont{A.~V.} \bibnamefont{Chubukov}} \bibnamefont{and}
  \bibinfo{author}{\bibfnamefont{M.~R.} \bibnamefont{Norman}},
  \bibinfo{journal}{Phys. Rev. B} \textbf{\bibinfo{volume}{70}},
  \bibinfo{pages}{174505} (\bibinfo{year}{2004}).

\bibitem[{\citenamefont{Sandvik et~al.}(2004)\citenamefont{Sandvik, Scalapino,
  and Bickers}}]{Sandvik-2004}
\bibinfo{author}{\bibfnamefont{A.~W.} \bibnamefont{Sandvik}},
  \bibinfo{author}{\bibfnamefont{D.~J.} \bibnamefont{Scalapino}},
  \bibnamefont{and} \bibinfo{author}{\bibfnamefont{N.~E.}
  \bibnamefont{Bickers}}, \bibinfo{journal}{Phys. Rev. B}
  \textbf{\bibinfo{volume}{69}}, \bibinfo{pages}{094523}
  (\bibinfo{year}{2004}).

\bibitem[{\citenamefont{Devereaux et~al.}(2004)\citenamefont{Devereaux, Cuk,
  Shen, and Nagaosa}}]{Devereaux-2004}
\bibinfo{author}{\bibfnamefont{T.~P.} \bibnamefont{Devereaux}},
  \bibinfo{author}{\bibfnamefont{T.}~\bibnamefont{Cuk}},
  \bibinfo{author}{\bibfnamefont{Z.-X.} \bibnamefont{Shen}}, \bibnamefont{and}
  \bibinfo{author}{\bibfnamefont{N.}~\bibnamefont{Nagaosa}},
  \bibinfo{journal}{Phys. Rev. Lett.} \textbf{\bibinfo{volume}{93}},
  \bibinfo{pages}{117004} (\bibinfo{year}{2004});
  \bibinfo{author}{\bibfnamefont{T.}~\bibnamefont{Cuk}},
  \bibinfo{author}{\bibfnamefont{D.~H.}~\bibnamefont{Lu}},
  \bibinfo{author}{\bibfnamefont{X.~Z.}~\bibnamefont{Zhou}},
  \bibinfo{author}{\bibfnamefont{Z.-X.} \bibnamefont{Shen}},
  \bibinfo{author}{\bibfnamefont{T.~P.} \bibnamefont{Devereaux}}, 
  \bibnamefont{and}
  \bibinfo{author}{\bibfnamefont{N.}~\bibnamefont{Nagaosa}},
  \bibinfo{journal}{Phys. Status Solidi B} \textbf{\bibinfo{volume}{242}},
  \bibinfo{pages}{11} (\bibinfo{year}{2005}).

\bibitem[{\citenamefont{Zhu et~al.}(2006)\citenamefont{Zhu, Balatsky,
  Devereaux, Si, Lee, McElroy, and Davis}}]{Zhu-2006}
\bibinfo{author}{\bibfnamefont{J.-X.} \bibnamefont{Zhu}},
  \bibinfo{author}{\bibfnamefont{A.~V.} \bibnamefont{Balatsky}},
  \bibinfo{author}{\bibfnamefont{T.~P.} \bibnamefont{Devereaux}},
  \bibinfo{author}{\bibfnamefont{Q.}~\bibnamefont{Si}},
  \bibinfo{author}{\bibfnamefont{J.}~\bibnamefont{Lee}},
  \bibinfo{author}{\bibfnamefont{K.}~\bibnamefont{McElroy}}, \bibnamefont{and}
  \bibinfo{author}{\bibfnamefont{J.~C.} \bibnamefont{Davis}},
  \bibinfo{journal}{Phys. Rev. B} \textbf{\bibinfo{volume}{73}},
  \bibinfo{pages}{014511} (\bibinfo{year}{2006}).

\bibitem[{\citenamefont{Onufrieva and Pfeuty}(2009)}]{Onufrieva-2009}
\bibinfo{author}{\bibfnamefont{F.}~\bibnamefont{Onufrieva}} \bibnamefont{and}
  \bibinfo{author}{\bibfnamefont{P.}~\bibnamefont{Pfeuty}},
  \bibinfo{journal}{Phys. Rev. Lett.} \textbf{\bibinfo{volume}{102}},
  \bibinfo{pages}{207003} (\bibinfo{year}{2009}).

\bibitem[{\citenamefont{Sidis et~al.}(2004)\citenamefont{Sidis, Pailh{\`e}s,
  Keimer, Bourges, Ulrich, and Regnault}}]{Sidis-2004}
\bibinfo{author}{\bibfnamefont{Y.}~\bibnamefont{Sidis}},
  \bibinfo{author}{\bibfnamefont{S.}~\bibnamefont{Pailh{\`e}s}},
  \bibinfo{author}{\bibfnamefont{B.}~\bibnamefont{Keimer}},
  \bibinfo{author}{\bibfnamefont{P.}~\bibnamefont{Bourges}},
  \bibinfo{author}{\bibfnamefont{C.}~\bibnamefont{Ulrich}}, \bibnamefont{and}
  \bibinfo{author}{\bibfnamefont{L.~P.} \bibnamefont{Regnault}},
  \bibinfo{journal}{Phys. Status Solidi B} \textbf{\bibinfo{volume}{241}},
  \bibinfo{pages}{1204} (\bibinfo{year}{2004}).

\bibitem[{\citenamefont{Jenkins et~al.}(2009)\citenamefont{Jenkins, Fasano,
  Berthod, Maggio-Aprile, Piriou, Giannini, Hoogenboom, Hess, Cren, and
  Fischer}}]{Jenkins-2009}
\bibinfo{author}{\bibfnamefont{N.}~\bibnamefont{Jenkins}},
  \bibinfo{author}{\bibfnamefont{Y.}~\bibnamefont{Fasano}},
  \bibinfo{author}{\bibfnamefont{C.}~\bibnamefont{Berthod}},
  \bibinfo{author}{\bibfnamefont{I.}~\bibnamefont{Maggio-Aprile}},
  \bibinfo{author}{\bibfnamefont{A.}~\bibnamefont{Piriou}},
  \bibinfo{author}{\bibfnamefont{E.}~\bibnamefont{Giannini}},
  \bibinfo{author}{\bibfnamefont{B.~W.} \bibnamefont{Hoogenboom}},
  \bibinfo{author}{\bibfnamefont{C.}~\bibnamefont{Hess}},
  \bibinfo{author}{\bibfnamefont{T.}~\bibnamefont{Cren}}, \bibnamefont{and}
  \bibinfo{author}{\bibfnamefont{{\O}.}~\bibnamefont{Fischer}},
  \bibinfo{journal}{Phys. Rev. Lett.} \textbf{\bibinfo{volume}{103}},
  \bibinfo{pages}{227001} (\bibinfo{year}{2009}).

\bibitem[{\citenamefont{Tersoff and Hamann}(1983)}]{Tersoff-1983}
\bibinfo{author}{\bibfnamefont{J.}~\bibnamefont{Tersoff}} \bibnamefont{and}
  \bibinfo{author}{\bibfnamefont{D.~R.} \bibnamefont{Hamann}},
  \bibinfo{journal}{Phys. Rev. Lett.} \textbf{\bibinfo{volume}{50}},
  \bibinfo{pages}{1998} (\bibinfo{year}{1983}).

\bibitem[{\citenamefont{Chen}(1990)}]{Chen-1990b}
\bibinfo{author}{\bibfnamefont{C.~J.} \bibnamefont{Chen}},
  \bibinfo{journal}{Phys. Rev. B} \textbf{\bibinfo{volume}{42}},
  \bibinfo{pages}{8841} (\bibinfo{year}{1990}).

\bibitem{self-consistency}
In electron-phonon models, the Migdal argument allows to neglect vertex
corrections and obtain an accurate self-consistent theory by replacing
$\hat{G}_0$ in the self-energy by the full $\hat{G}$
(Refs~\onlinecite{Migdal-1958} and \onlinecite{McMillan-1969}). As there is no
such theorem for spin fluctuations, the self-consistent model $\chi_s\hat{G}$ is
not necessarily better (\textit{i.e.}, closer to the exact result including all
vertex corrections) than the lowest-order model $\chi_s\hat{G}_0$. This issue
has been pointed out in the case of the Hubbard model
(Ref.~\onlinecite{Vilk-1997}), and is also well known in the context of the
$GW$ approximation for the long-range Coulomb interaction, where the
self-consistent version often gives poorer results than the non self-consistent
one (Ref.~\onlinecite{Hedin-1999}). The situation is further complicated if one
uses a phenomenological spin susceptibility in Eq.~(\ref{eq:Sigma}), rather than
the spin susceptibility calculated using a consistent approximation. No
systematic inclusion of vertex corrections is possible in this latter case.
Hence the justification of Eq.~(\ref{eq:Sigma}) and of the corresponding
self-consistent model (Ref.~\onlinecite{Onufrieva-2009}) resides in their
ability to properly describe experiments.

\bibitem[{\citenamefont{Migdal}(1958)}]{Migdal-1958}
\bibinfo{author}{\bibfnamefont{A.~B.} \bibnamefont{Migdal}},
  \bibinfo{journal}{Sov. Phys. JETP} \textbf{\bibinfo{volume}{7}},
  \bibinfo{pages}{996} (\bibinfo{year}{1958}).

\bibitem[{\citenamefont{McMillan and Rowell}(1969)}]{McMillan-1969}
\bibinfo{author}{\bibfnamefont{W.~L.} \bibnamefont{McMillan}} \bibnamefont{and}
  \bibinfo{author}{\bibfnamefont{J.~M.} \bibnamefont{Rowell}}, in
  \emph{\bibinfo{booktitle}{Superconductivity}}, edited by
  \bibinfo{editor}{\bibfnamefont{R.~D.} \bibnamefont{Parks}}
  (\bibinfo{publisher}{Dekker}, \bibinfo{address}{New York},
  \bibinfo{year}{1969}), Vol.~\bibinfo{volume}{1}, p. \bibinfo{pages}{561}.

\bibitem[{\citenamefont{Vilk and Tremblay}(1997)}]{Vilk-1997}
\bibinfo{author}{\bibfnamefont{Y.~M.} \bibnamefont{Vilk}} \bibnamefont{and}
  \bibinfo{author}{\bibfnamefont{A.-M.~S.} \bibnamefont{Tremblay}},
  \bibinfo{journal}{J. Phys. I} \textbf{\bibinfo{volume}{7}},
  \bibinfo{pages}{1309} (\bibinfo{year}{1997}).

\bibitem[{\citenamefont{Hedin}(1999)}]{Hedin-1999}
\bibinfo{author}{\bibfnamefont{L.}~\bibnamefont{Hedin}}, \bibinfo{journal}{J.
  Phys.: Condens. Matter} \textbf{\bibinfo{volume}{11}}, \bibinfo{pages}{R489}
  (\bibinfo{year}{1999}).

\bibitem[{\citenamefont{Rajagopal and Jha}(1996)}]{Rajagopal-1996}
\bibinfo{author}{\bibfnamefont{A.~K.} \bibnamefont{Rajagopal}} \bibnamefont{and}
  \bibinfo{author}{\bibfnamefont{S. S.} \bibnamefont{Jha}},
  \bibinfo{journal}{Phys. Rev. B} \textbf{\bibinfo{volume}{54}},
  \bibinfo{pages}{4331} (\bibinfo{year}{1996});
  \bibinfo{author}{\bibfnamefont{S. S.} \bibnamefont{Jha}} \bibnamefont{and}
  \bibinfo{author}{\bibfnamefont{A.~K.} \bibnamefont{Rajagopal}},
  \bibinfo{journal}{Phys. Rev. B} \textbf{\bibinfo{volume}{55}},
  \bibinfo{pages}{15248} (\bibinfo{year}{1997}).

\bibitem[{\citenamefont{Pairor}(2005)}]{Pairor-2005}
\bibinfo{author}{\bibfnamefont{P.} \bibnamefont{Pairor}},
  \bibinfo{journal}{Phys. Rev. B} \textbf{\bibinfo{volume}{72}},
  \bibinfo{pages}{174519} (\bibinfo{year}{2005}).

\bibitem[{\citenamefont{Won and Maki}(1994)}]{Won-1994}
\bibinfo{author}{\bibfnamefont{H.}~\bibnamefont{Won}} \bibnamefont{and}
  \bibinfo{author}{\bibfnamefont{K.}~\bibnamefont{Maki}},
  \bibinfo{journal}{Phys. Rev. B} \textbf{\bibinfo{volume}{49}},
  \bibinfo{pages}{1397} (\bibinfo{year}{1994}).

\bibitem[{\citenamefont{Annett}(1990)}]{Annett-1990}
\bibinfo{author}{\bibfnamefont{J.~F.} \bibnamefont{Annett}},
  \bibinfo{journal}{Adv. Phys.} \textbf{\bibinfo{volume}{39}},
  \bibinfo{pages}{83} (\bibinfo{year}{1990}).

\bibitem{endnote1}
Just the opposite happens in electron-phonon models, which are pair breaking in
the $d$-wave channel. In electron-phonon models, the self-energy
Eq.~(\ref{eq:Sigma}) is replaced by, schematically, $\hat{\Sigma}\propto
-g^2D\hat{\tau}_3\hat{G}_0\hat{\tau}_3$ with $g$ the electron-phonon coupling
and $D$ the phonon propagator (Ref.~\onlinecite{Schrieffer-1964b}). The
$\hat{\tau}_3$ matrices appear because phonons couple to the charge
density---while spin fluctuations couple to the spin density, and they change
the sign of the $\hat{\Sigma}_{12}$ component controlling the gap
renormalization.

\bibitem[{\citenamefont{Lee et~al.}(2006)\citenamefont{Lee, Fujita, McElroy,
  Slezak, Wang, Aiura, Bando, Ishikado, Masui, Zhu et~al.}}]{Lee-2006}
\bibinfo{author}{\bibfnamefont{J.}~\bibnamefont{Lee}},
  \bibinfo{author}{\bibfnamefont{K.}~\bibnamefont{Fujita}},
  \bibinfo{author}{\bibfnamefont{K.}~\bibnamefont{McElroy}},
  \bibinfo{author}{\bibfnamefont{J.~A.} \bibnamefont{Slezak}},
  \bibinfo{author}{\bibfnamefont{M.}~\bibnamefont{Wang}},
  \bibinfo{author}{\bibfnamefont{Y.}~\bibnamefont{Aiura}},
  \bibinfo{author}{\bibfnamefont{H.}~\bibnamefont{Bando}},
  \bibinfo{author}{\bibfnamefont{M.}~\bibnamefont{Ishikado}},
  \bibinfo{author}{\bibfnamefont{T.}~\bibnamefont{Masui}},
  \bibinfo{author}{\bibfnamefont{J.~X.} \bibnamefont{Zhu}},
  \bibinfo{author}{\bibfnamefont{A.~V.} \bibnamefont{Balatsky}},
  \bibinfo{author}{\bibfnamefont{H.} \bibnamefont{Eisaki}},
  \bibinfo{author}{\bibfnamefont{S.} \bibnamefont{Uchida}},
  \bibnamefont{and}
  \bibinfo{author}{\bibfnamefont{J.~C.} \bibnamefont{Davis}},
  \bibinfo{journal}{Nature}
  \textbf{\bibinfo{volume}{442}}, \bibinfo{pages}{546} (\bibinfo{year}{2006}).

\bibitem[{\citenamefont{Zhao}(2007)}]{Zhao-2007}
\bibinfo{author}{\bibfnamefont{Guo-meng} \bibnamefont{Zhao}},
  \bibinfo{journal}{Phys. Rev. B} \textbf{\bibinfo{volume}{75}},
  \bibinfo{pages}{214507} (\bibinfo{year}{2007}).

\bibitem[{\citenamefont{Bourges et~al.}(2000)\citenamefont{Bourges, Keimer,
  Regnault, and Sidis}}]{Bourges-2000a}
\bibinfo{author}{\bibfnamefont{P.}~\bibnamefont{Bourges}},
  \bibinfo{author}{\bibfnamefont{B.}~\bibnamefont{Keimer}},
  \bibinfo{author}{\bibfnamefont{L.~P.} \bibnamefont{Regnault}},
  \bibnamefont{and} \bibinfo{author}{\bibfnamefont{Y.}~\bibnamefont{Sidis}},
  \bibinfo{journal}{J. Supercond.} \textbf{\bibinfo{volume}{13}},
  \bibinfo{pages}{735} (\bibinfo{year}{2000}).

\bibitem[{\citenamefont{Regnault et~al.}(1995)\citenamefont{Regnault, Bourges,
  Burlet, Henry, Rossat-Mignod, Sidis, and Vettier}}]{Regnault-1995}
\bibinfo{author}{\bibfnamefont{L.~P.} \bibnamefont{Regnault}},
  \bibinfo{author}{\bibfnamefont{P.}~\bibnamefont{Bourges}},
  \bibinfo{author}{\bibfnamefont{P.}~\bibnamefont{Burlet}},
  \bibinfo{author}{\bibfnamefont{J.~Y.} \bibnamefont{Henry}},
  \bibinfo{author}{\bibfnamefont{J.}~\bibnamefont{Rossat-Mignod}},
  \bibinfo{author}{\bibfnamefont{Y.}~\bibnamefont{Sidis}}, \bibnamefont{and}
  \bibinfo{author}{\bibfnamefont{C.}~\bibnamefont{Vettier}},
  \bibinfo{journal}{Physica B} \textbf{\bibinfo{volume}{213-214}},
  \bibinfo{pages}{48} (\bibinfo{year}{1995}).

\bibitem{endnote2}
The 200~K data in Fig.~7 of Ref.~\onlinecite{Regnault-1995} can be well fitted
to the Lorentzian form used for the spin susceptibility in Eq.~(\ref{eq:Sigma}),
namely, $A\{(\Gamma_{\text{sf}}/\pi)/[(\omega-\Omega_{\text{sf}})^2+
\Gamma_{\text{sf}}^2]-(\Gamma_{\text{sf}}/\pi)/[(\omega+\Omega_
{\text{sf}})^2+\Gamma_{\text{sf}}^2]\}$, with the parameters
$\Omega_{\text{sf}}=25$~meV, $\Gamma_{\text{sf}}=14$~meV, and $A=5519$. On the
other hand, the momentum width of the spin response was found to be temperature
independent. To model the normal state of Bi-2223, we may therefore keep $\Delta
q$ fixed and increase $\Gamma_{\text{sr}}$ to the value of
$\Gamma_{\text{sf}}=14$~meV. In the absence of more detailed information, we do
not change the value of $\Omega_{\text{sr}}$, which now plays the role of the
broad maximum in the magnetic response. These new parameters lead to a reduced
effective mass $m^*/m=1.71$ in the normal state as compared to the
superconducting state.

\bibitem[{\citenamefont{Romano et~al.}(2006)\citenamefont{Romano, Ozyuzer,
  Yusof, Kurter, and Zasadzinski}}]{Romano-2006}
\bibinfo{author}{\bibfnamefont{P.}~\bibnamefont{Romano}},
  \bibinfo{author}{\bibfnamefont{L.}~\bibnamefont{Ozyuzer}},
  \bibinfo{author}{\bibfnamefont{Z.}~\bibnamefont{Yusof}},
  \bibinfo{author}{\bibfnamefont{C.}~\bibnamefont{Kurter}}, \bibnamefont{and}
  \bibinfo{author}{\bibfnamefont{J.~F.} \bibnamefont{Zasadzinski}},
  \bibinfo{journal}{Phys. Rev. B} \textbf{\bibinfo{volume}{73}},
  \bibinfo{pages}{092514} (\bibinfo{year}{2006}).

\bibitem[{\citenamefont{Boyer et~al.}(2007)\citenamefont{Boyer, Wise, Kamalesh,
  Yi, Kondo, Takeuchi, Ikuta, and Hudson}}]{Boyer-2007}
\bibinfo{author}{\bibfnamefont{M.~C.} \bibnamefont{Boyer}},
  \bibinfo{author}{\bibfnamefont{W.~D.} \bibnamefont{Wise}},
  \bibinfo{author}{\bibfnamefont{C.}~\bibnamefont{Kamalesh}},
  \bibinfo{author}{\bibfnamefont{M.}~\bibnamefont{Yi}},
  \bibinfo{author}{\bibfnamefont{T.}~\bibnamefont{Kondo}},
  \bibinfo{author}{\bibfnamefont{T.}~\bibnamefont{Takeuchi}},
  \bibinfo{author}{\bibfnamefont{H.}~\bibnamefont{Ikuta}}, \bibnamefont{and}
  \bibinfo{author}{\bibfnamefont{E.~W.} \bibnamefont{Hudson}},
  \bibinfo{journal}{Nat. Phys.} \textbf{\bibinfo{volume}{3}},
  \bibinfo{pages}{802} (\bibinfo{year}{2007}).

\bibitem[{\citenamefont{Pasupathy et~al.}(2008)\citenamefont{Pasupathy, Pushp,
  Gomes, Parker, Wen, Xu, Gu, Ono, Ando, and Yazdani}}]{Pasupathy-2008}
\bibinfo{author}{\bibfnamefont{A.~N.} \bibnamefont{Pasupathy}},
  \bibinfo{author}{\bibfnamefont{A.}~\bibnamefont{Pushp}},
  \bibinfo{author}{\bibfnamefont{K.~K.} \bibnamefont{Gomes}},
  \bibinfo{author}{\bibfnamefont{C.~V.} \bibnamefont{Parker}},
  \bibinfo{author}{\bibfnamefont{J.}~\bibnamefont{Wen}},
  \bibinfo{author}{\bibfnamefont{Z.}~\bibnamefont{Xu}},
  \bibinfo{author}{\bibfnamefont{G.}~\bibnamefont{Gu}},
  \bibinfo{author}{\bibfnamefont{S.}~\bibnamefont{Ono}},
  \bibinfo{author}{\bibfnamefont{Y.}~\bibnamefont{Ando}}, \bibnamefont{and}
  \bibinfo{author}{\bibfnamefont{A.}~\bibnamefont{Yazdani}},
  \bibinfo{journal}{Science} \textbf{\bibinfo{volume}{320}},
  \bibinfo{pages}{196} (\bibinfo{year}{2008}).


\end{thebibliography}
\end{document}